# Cluster folding analysis for the $^{6,7}$Li and $^{20}$Ne + $^{24}$Mg nuclear systems

Sh. Hamada[*]

[*]Faculty of Science, Tanta University, Tanta, Egypt
sh.m.hamada@science.tanta.edu.eg

**Abstract:** The experimentally available angular distributions (ADs) for $^{6,7}$Li and $^{20}$Ne ions elastically scattered from a $^{24}$Mg target are reanalyzed using various nuclear potentials based on phenomenological and microscopic approaches. The ADs for the $^{6}$Li + $^{24}$Mg system at energies $E_{lab}$ = 20, 88, and 240 MeV are considered, as are the ADs for the $^{7}$Li + $^{24}$Mg system at $E_{lab}$ = 20, 34, and 88 MeV, and the ADs for the $^{20}$Ne + $^{24}$Mg systems at $E_{lab}$ = 50, 60, 80, 90, and 100 MeV. Special emphasis is placed on the ($d + \alpha$) and ($t + \alpha$) cluster structures of $^{6}$Li and $^{7}$Li weakly bound projectiles, which appear at relatively low energies of 1.4737 and 2.468 MeV, respectively. Furthermore, the $^{20}$Ne stable nucleus is a candidate to reveal the $\alpha + {}^{16}$O cluster structure at an energy of 4.73 MeV. These systems are analyzed within the framework of the cluster folding model by taking into consideration the aforementioned cluster nature of the ($^{6}$Li, $^{7}$Li, and $^{20}$Ne) projectiles. The adopted approaches fairly reproduced the experimental ADs.

*Keywords:* Elastic scattering; Optical model potential; Sao Paulo potential; Cluster folding model

*PACS number(s)*: "24.10.Eq, 21.10.Jx, 25.70.Hi, 21.60.Cs"

## I. Introduction

The nucleus is traditionally described as having a homogenous distribution of neutrons and protons. However, it was well-known even from the beginning of nuclear research that aggregates of nucleons (nuclear clustering) could lead to a better understanding of nuclei structure as well as interactions mechanisms. In the 1950s, Morinaga [1] proposed an extreme prediction; the *α*-particles should be able to arrange themselves linearly. The idea is that, the cluster structure should not be appearing in the nucleus ground state, but it starts to appear with increasing the internal energy of the nucleus. So, a nucleus must have a certain amount of energy to develop a cluster structure. In another words, the cluster structure is expected to appear near and slightly below the cluster decay threshold energy. Ikeda et al. [2] expected that cluster structure would be most clear at excitation energy ($E_x$) corresponded with a specific decay threshold. As a result, in its ground state, the cluster structure of $^{8}$Be as $\alpha + \alpha$ can be observed (which decays into two *α* -particles within $10^{-16}$ seconds), and the three *α*-cluster structure in $^{12}$C is expected to appear at threshold energy ~ 7.27 MeV. The Hoyle state in $^{12}$C at $E_x$ = 7.65 MeV which is of special interest for nuclear astrophysicists, is believed to have a well-developed three *α*-structure.

Several experimental evidences for the concept clusterization in light nuclei are presented in the work of M. Freer [3]. The simplest example is the two *α*-particle systems in $^{8}$Be. The *α*-particle binding energy is so large (~28 MeV), so $^{6}$Li and $^{7}$Li nuclei could display the cluster structures $d + \alpha +$ and $t + \alpha$, respectively. The Hoyle-state in $^{12}$C at $E_x$ = 7.65 MeV is a perfect cluster state. Hoyle [4] predicted the state to explain the amount of carbon in the universe, and Cook [5] detected this state at nearly the same energy. In stellar environments, carbon is formed into the triple-alpha process, in which two *α* -particles join momentarily to create $^{8}$Be, and then a third *α* -particle is captured before the system decays. This mostly occurs through the second excited state ($0^+$) at 7.65 MeV, which then radioactively decay at 4.43 MeV to the $^{12}$C ground state. The radius of this state is known to be extremely large, allowing the *α* -particles to preserve their quasi-free characteristics.

In the field of nuclear physics, elastic scattering processes induced by stable atomic nuclei are still a rich area of research. Depending on the structure of the two colliding nuclei and on the projectile's energy, the elastic scattering angular distributions (ADs) can display various features. A Fresnel oscillatory diffraction pattern may develop when the ADs are plotted as a ratio to Rutherford cross sections for stable projectiles at energies near to the Coulomb barrier. The Fresnel peak, also known as the Coulomb-nuclear interference peak or Coulomb rainbow, is caused by the interference between partial waves refracted by the Coulomb and the short range nuclear potentials. At higher energies, the Coulomb force for light projectiles decreases, and the diffractive pattern changes from Fresnel to Fraunhofer [6]. While the ADs for processes induced by stable and tightly bound nuclei display one of the standard diffraction patterns, the ADs for those induced by weakly bound nuclei such as $^{6,7}$Li deviate significantly from the oscillatory pattern. Low binding energies between the valence particle(s) and the core characterize these nuclei, resulting in some decoupling during the collision. This effect leads to the appearance of non-elastic processes, even over long distances, and finally causes the Fresnel peak to be dampened or completely disappear.

The $\alpha + {}^{16}$O model of $^{20}$Ne nucleus has recently been applied in investigating different nuclear systems induced by $^{20}$Ne or in systems treated in inverse kinematics where the $^{20}$Ne nucleus is the bombarded target such as $^{20}$Ne + $^{20}$Ne [7], $^{20}$Ne + $^{16}$O [8,9], $\alpha + {}^{20}$Ne [10] and p + $^{20}$Ne [11] systems. Consequently, it is interesting to apply such model in investigating the $^{20}$Ne + $^{24}$Mg system, which is less experimentally studied and has not been investigated from the microscopic point of view. This work aims to study the probable $\alpha + {}^{16}$O cluster structure for the $^{20}$Ne nucleus. For to this purpose, the previously measured elastic scattering ADs for $^{20}$Ne in the field of $^{24}$Mg are reanalyzed for the first time from the microscopic point of view. In addition to the $^{20}$Ne + $^{24}$Mg system, the $^{6,7}$Li + $^{24}$Mg systems are also reanalyzed using several potentials, starting from the simplest and widely used optical model potential (OMP), the double-folded Sao Paulo potential (SPP) and finally by applying the more sophisticated cluster folding model (CFM).

The manuscript is organized as follow: Sec. II presents the implemented theoretical methods. Results and discussion are given in Sec. III. The summary is presented in Sec. IV.

**II. Implemented theoretical methods**

In the present work, the experimentally available ADs for $^{6}$Li, $^{7}$Li, and $^{20}$Ne elastically scattered from a $^{24}$Mg target are subjected to detailed analysis using various nuclear potentials. The ADs for the $^{6}$Li + $^{24}$Mg system at energies $E_{lab}$ = 20, 88, and 240 MeV [12-14], the $^{7}$Li + $^{24}$Mg system at $E_{lab}$ = 20, 34, and 88 MeV [12, 15, 16], and the $^{20}$Ne + $^{24}$Mg system at $E_{lab}$ = 50–100 MeV [17] are considered. Although the elastic scattering ADs for the $^{20}$Ne + $^{24}$Mg system at $E_{lab}$= 50–100 MeV were measured many years ago [17], these data were never thoroughly examined using microscopic approaches. As a first step, we analyze the considered data utilizing the conventional optical model (OM) of two varying parameters (real and imaginary potential depths) and fixed geometrical parameters (radius and diffuseness). Then, these data are investigated using the Sao Paulo potential (SPP), which considers the density distributions of the interacting nuclei. Finally, the full microscopic CFM is used to reproduce the $^{6}$Li + $^{24}$Mg and $^{7}$Li + $^{24}$Mg ADs utilizing the ($d + \alpha$) and ($t + \alpha$) cluster structures for $^{6}$Li and $^{7}$Li, respectively, as well as to test the validity of reproducing the $^{20}$Ne + $^{24}$Mg data by applying the $\alpha + {}^{16}$O model for $^{20}$Ne, which demonstrated significant success in describing the experimental data for many nuclear systems induced by $^{20}$Ne [7-11].

### A. Optical Model potential (OMP)

Analysis of experimental data was first performed using the phenomenological OMP approach. The potential parameters were chosen to achieve the best possible agreement with the data. Data fitting was carried out in the full angular range. The Woods-Saxon (WS) form factor shape was chosen to express both the real and imaginary parts of the potential as shown in Eq. (1).

$$U(R) = V_C - V_0 \left[1 + \exp\left(\frac{r - R_V}{a_V}\right)\right]^{-1} - iW_0 \left[1 + \exp\left(\frac{r - R_W}{a_W}\right)\right]^{-1} \tag{1}$$

The first term is the Coulomb potential of a uniform charged sphere with radius:

$R_i = r_i(A_P^{1/3} + A_T^{1/3})$, $i = V, W, C$ "the used form in the analysis of $^{20}$Ne + $^{24}$Mg system", and $R_i = r_i(A_T^{1/3})$, $i = V, W, C$ "the used form in the analysis of $^{6,7}$Li + $^{24}$Mg systems". For the $^{20}$Ne + $^{24}$Mg system, the OMP analysis started with the same geometrical parameters as Belery et al. [17]. The radius parameter $r_V$ and $r_W$ for the real and imaginary parts of the nuclear potential are fixed to 1.25 fm, and the diffuseness $a_V$ and $a_W$ for the real and imaginary parts are fixed to 0.65 fm, allowing the other two parameters $V_0$ and $W_0$ (real and imaginary potential depths) to vary till the best agreement to the experimental data is achieved through minimizing the $\chi^2$ value, which gives the deviation between measurements and calculations, defined by,

$$\chi^2 = \frac{1}{N}\sum_{i=1}^{N}\left(\frac{\sigma(\theta_i)^{cal} - \sigma(\theta_i)^{exp}}{\Delta\sigma(\theta_i)}\right)^2, \tag{2}$$

The quantities $N$, $\sigma(\theta_i)^{cal}$, $\sigma(\theta_i)^{exp}$ and $\Delta\sigma(\theta_i)$ represent the number of data points, calculated and experimental differential cross sections, and the data relative uncertainty, respectively. The theoretical calculations as well as searching for the best potential parameters are done using the **FRESCO** and **SFRESCO** codes [18].

In accordance with previous studies concerning $^{6,7}$Li + $^{24}$Mg systems, the starting geometrical parameters used in OMP analysis for the $^6$Li + $^{24}$Mg system are taken from Ref. [13]. For $^7$Li+$^{24}$Mg, the global $^7$Li potential of Cook [19] is applied. It worth to mention that, although the experimental data for the $^{20}$Ne + $^{24}$Mg system are limited to 80° and did not extend to larger angles to check the expected contribution of α-cluster transfer between $^{20}$Ne and $^{24}$Mg nuclei, the data still could be of special interest to investigate the appearance of the α + $^{16}$O structure in $^{20}$Ne ground state.

### B. Sao Paulo potential (SPP)

In order to avoid parameter ambiguities that might be attributed to OM calculations, the more microscopic SPP is employed to generate the real part of the potential by folding the target ($\rho_t$) and the projectile ($\rho_p$) densities obtained from the Dirac-Hartree-Bogoliubov (DHB) model [20] with an effective potential expressed as:

$$V_F(R) = \iint \rho_P(r_P)\rho_T(r_T)V_0\delta((|\vec{s}|))d^3r_P\, d^3r_T, \quad \vec{s} = \vec{R} - \vec{r}_P + \vec{r}_T \tag{3}$$

The new version of the Sao Paulo potential (SPP2) was implemented using the **REGINA** code [21] with nuclear densities obtained from the DHB Model. The prepared SPPs for the $^6$Li + $^{24}$Mg system at $E_{lab}$ = 20, 88, and 240 MeV, the $^7$Li + $^{24}$Mg system at $E_{lab}$ = 20, 34, and 88 MeV, and the $^{20}$Ne + $^{24}$Mg system at $E_{lab}$ = 50, 60, 80, 90, and 100 MeV using Eq. (3) are presented in Fig.

1. As shown in Fig. 1, the differences in potential with respect to energy are obvious in the region at R < 6 fm for $^6$Li+$^{24}$Mg and $^7$Li+$^{24}$Mg systems, "see Fig. 1a and 1b". For the $^{20}$Ne + $^{24}$Mg system, these differences are clear in the internal region at R < 2 fm.

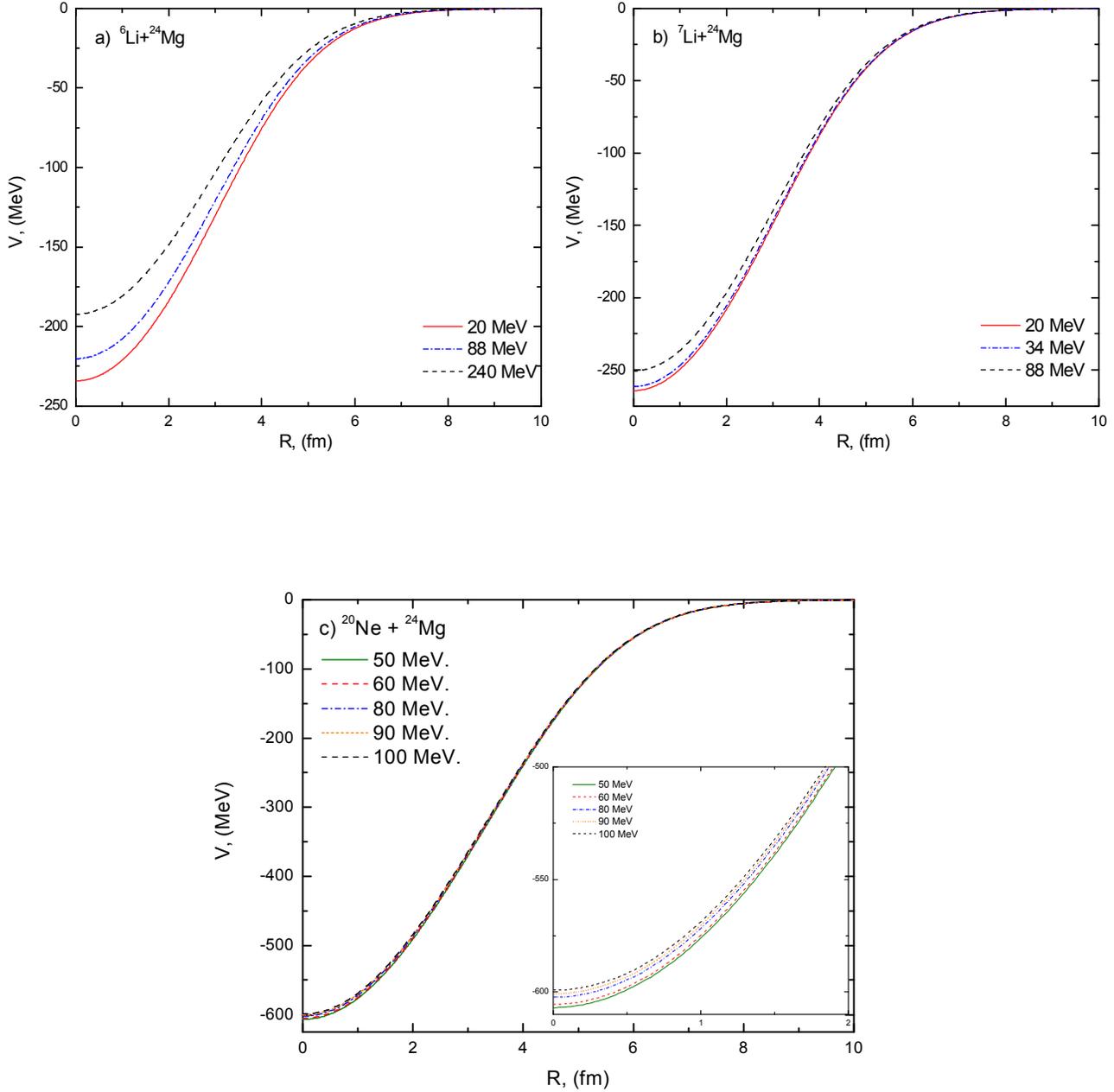

**Fig. 1:** The generated real SPP at for a) $^6$Li + $^{24}$Mg at $E_{lab}$ = 20, 88, and 240 MeV, b) $^7$Li + $^{24}$Mg at $E_{lab}$ = 20, 34, and 88 MeV, and c) $^{20}$Ne + $^{24}$Mg at $E_{lab}$ = 50, 60, 80, 90 and 100 MeV. The inner panel in Fig. 2c is just to clarify the differences in potential with respect to energy, the differences are clear in the internal region at R < 2 fm.

## C. Cluster Folding Potential (CFP)

It is interesting to explore the possibility of reproducing the available experimental ADs for $^{6,7}$Li + $^{24}$Mg and $^{20}$Ne + $^{24}$Mg systems using CFM, motivated by the well-known $d + \alpha$ and $t + \alpha$ cluster structures for $^{6}$Li and $^{7}$Li, as well as the expected $\alpha + ^{16}$O cluster structure for $^{20}$Ne. Within the framework of CFM, the real and imaginary parts of the potential were constructed based on the cluster folding (CF) approach. The real and imaginary CF potentials for $^{6,7}$Li + $^{24}$Mg and $^{20}$Ne + $^{24}$Mg systems are presented in Fig. 2. The following procedures are followed to create the CFPs:

***a)*** Based on the $\alpha + ^{24}$Mg and $d + ^{24}$Mg potentials, the real and imaginary cluster folding parts of the $^{6}$Li + $^{24}$Mg potential are defined as follows:

$$V^{CF}(\mathbf{R}) = \int \left[ V_{\alpha-^{24}\text{Mg}}\left(\mathbf{R}-\frac{1}{3}\mathbf{r}\right) + V_{d-^{24}\text{Mg}}\left(\mathbf{R}+\frac{2}{3}\mathbf{r}\right) \right] |\chi_{\alpha d}(\mathbf{r})|^2 d\mathbf{r}, \tag{4}$$

$$W^{CF}(\mathbf{R}) = \int \left[ W_{\alpha-^{24}\text{Mg}}\left(\mathbf{R}-\frac{1}{3}\mathbf{r}\right) + W_{d-^{24}\text{Mg}}\left(\mathbf{R}+\frac{2}{3}\mathbf{r}\right) \right] |\chi_{\alpha d}(\mathbf{r})|^2 d\mathbf{r}, \tag{5}$$

The required parameters to create the $^{6}$Li + $^{24}$Mg CFP are: the optimal potentials for $d + ^{24}$Mg and $\alpha + ^{24}$Mg channels at appropriate energies $E_d \approx 1/3E_{\text{Li}}$ and $E_\alpha \approx 2/3E_{\text{Li}}$ taken from Refs. [22, 23]. These potentials, in addition to the $\chi_{\alpha d}(\mathbf{r})$ intercluster wave function which describes the relative motion of $\alpha$ and $d$ in the ground state of $^{6}$Li were used to generate the real and imaginary CFPs expressed in Eqs. (4 and 5) as shown in Fig. 2a. The $\alpha$-$d$ bound state form factor represents a 2$S$ state in a real WS potential of radius and diffuseness equal 1.83 and 0.65 fm, respectively, and the depth is allowed to vary till the experimental binding energy of the cluster structure is obtained.

***b)*** Based on the $\alpha + ^{24}$Mg and $t + ^{24}$Mg potentials, the real and imaginary cluster folding parts of the $^{7}$Li + $^{24}$Mg potential are defined as follows:

$$V^{CF}(\mathbf{R}) = \int \left[ V_{\alpha-^{24}\text{Mg}}\left(\mathbf{R}-\frac{3}{7}\mathbf{r}\right) + V_{t-^{24}\text{Mg}}\left(\mathbf{R}+\frac{4}{7}\mathbf{r}\right) \right] |\chi_{\alpha-t}(\mathbf{r})|^2 d\mathbf{r}, \tag{6}$$

$$W^{CF}(\mathbf{R}) = \int \left[ W_{\alpha-^{24}\text{Mg}}\left(\mathbf{R}-\frac{3}{7}\mathbf{r}\right) + W_{t-^{24}\text{Mg}}\left(\mathbf{R}+\frac{4}{7}\mathbf{r}\right) \right] |\chi_{\alpha-t}(\mathbf{r})|^2 d\mathbf{r}, \tag{7}$$

where ($V_{\alpha-^{24}\text{Mg}}$ and $V_{t-^{24}\text{Mg}}$) and ($W_{\alpha-^{24}\text{Mg}}$ and $W_{t-^{24}\text{Mg}}$) are the real and imaginary parts of the potentials for $\alpha + ^{24}$Mg and $t + ^{24}$Mg channels, which reasonably fit the experimental data at energies $E_t \approx 3/7E_{\text{Li}}$ and $E_\alpha \approx 4/7E_{\text{Li}}$ taken from Refs. [24, 25]. $\chi_{\alpha-t}(\mathbf{r})$ is the intercluster wave function for the relative motion of $\alpha$ and $t$ in the ground state of $^{7}$Li, and $\mathbf{r}$ is the relative coordinate between the centers of mass of $\alpha$ and $t$. The $\alpha$-$t$ bound state form factor represents a 2$P_{3/2}$ state in a real WS potential of radius 1.83 fm, a diffuseness of 0.65 fm, and the depth is allowed to vary till the binding energy for the cluster structure is reached. The real and imaginary CFPs calculated according to Eqs. (6 and 7) are shown in Fig. 2b.

***c)*** The optimal potentials for $\alpha + ^{24}$Mg and $^{16}$O + $^{24}$Mg channels, which we shall refer to as V$_1$ for ($\alpha + ^{24}$Mg) and V$_2$ for ($^{16}$O + $^{24}$Mg), as well as the intercluster wave function $\chi_{\alpha-^{16}\text{O}}(\mathbf{r})$

which describes the α + $^{16}$O relative motion in $^{20}$Ne nucleus, are the essential requirements to generate the CFP for the $^{20}$Ne + $^{24}$Mg system. The real and imaginary CFPs for the $^{20}$Ne + $^{24}$Mg system are generated based on the α + $^{24}$Mg and $^{16}$O + $^{24}$Mg potentials as:

$$V^{CF}(\mathbf{R}) = \int \left[ V_{\alpha\,^{24}Mg}\left(\mathbf{R}-\frac{4}{5}\mathbf{r}\right) + V_{^{16}O\,^{24}Mg}\left(\mathbf{R}+\frac{1}{5}\mathbf{r}\right) \right] |\chi_{\alpha-^{16}O}(\mathbf{r})|^2 \, d\mathbf{r}, \tag{8}$$

$$W^{CF}(\mathbf{R}) = \int \left[ W_{\alpha\,^{24}Mg}\left(\mathbf{R}-\frac{4}{5}\mathbf{r}\right) + W_{^{16}O\,^{24}Mg}\left(\mathbf{R}+\frac{1}{5}\mathbf{r}\right) \right] |\chi_{\alpha-^{16}O}(\mathbf{r})|^2 \, d\mathbf{r}, \tag{9}$$

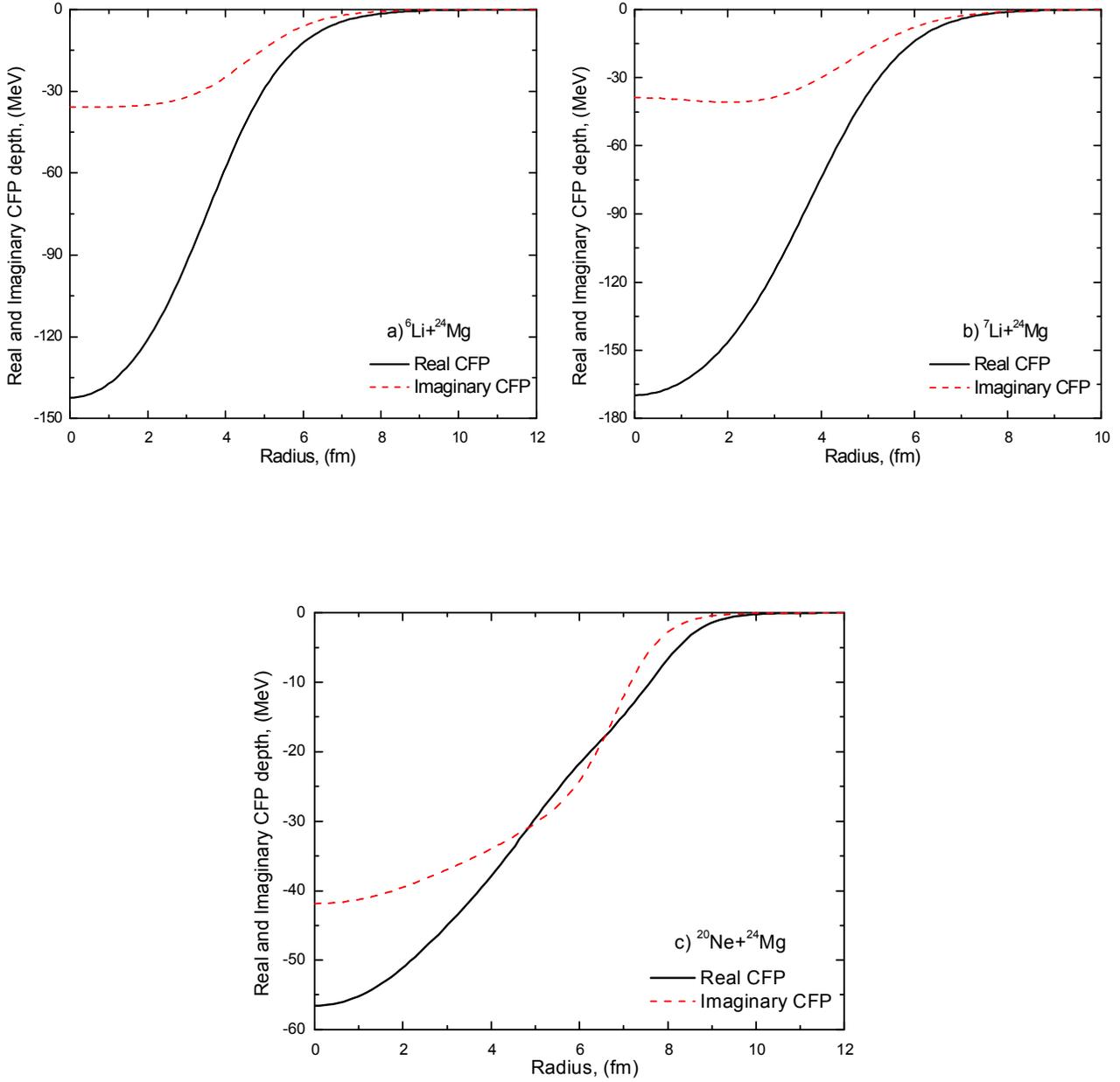

**Fig. 2**: The implemented real and imaginary CF potentials in data analysis.

where $V_{\alpha-^{24}Mg}$, $V_{^{16}O-^{24}Mg}$, $W_{\alpha-^{24}Mg}$, and $W_{^{16}O-^{24}Mg}$ are the phenomenological potentials for $\alpha$ + $^{24}$Mg and $^{16}$O + $^{24}$Mg channels taken from Refs. [26, 27] that adequately reproduce the experimental data at $E_\alpha \approx 1/5\ E_{Ne}$ and $E_{16O} \approx 4/5\ E_{Ne}$. The bound state form factor $\alpha$ + $^{16}$O represents a 5S state with binding potential taken from Ref. [28]. As the highest considered energy is 100 MeV, so the needed potentials are: $U_{\alpha-^{24}Mg}$ at $E_{lab}$ = 1/5 x 100 = 20 MeV and $U_{^{16}O-^{24}Mg}$ at $E_{lab}$ = 4/5 x 100 = 80 MeV, $U = (V+W)$ is the nuclear potential. The most suitable potentials found in literature, which could be used to generate the CFP for $^{20}$Ne + $^{24}$Mg are: $\alpha$ + $^{24}$Mg at $E_{lab}$ = 22.2 MeV [26] and $^{16}$O + $^{24}$Mg at $E_{lab}$ = 81 MeV [27]. The real and imaginary CFPs calculated according to Eqs. (8 and 9) are shown in Fig. 2c.

### III. Results and discussions
#### A. OM analysis for $^{6,7}$Li and $^{20}$Ne + $^{24}$Mg systems

The agreement between the $^6$Li, $^7$Li, $^{20}$Ne + $^{24}$Mg ADs and theoretical calculations within the framework of OMP is reasonably good in the full angular range as shown in Figs. 3–5, the optimal obtained potential parameters are listed in table I. Although the $^{6,7}$Li + $^{24}$Mg ADs were measured at limited energies ($E(^6$Li$)$ = 20, 88, and 240 MeV, and $E(^7$Li$)$ = 20, 34, and 88 MeV), the analysis showed an increase in imaginary potential depth at the lowest studied energy (20 MeV), which could be explained in terms of breakup threshold anomaly (BTA) observed in various systems induced by the weakly $^{6,7}$Li projectiles [29-36].

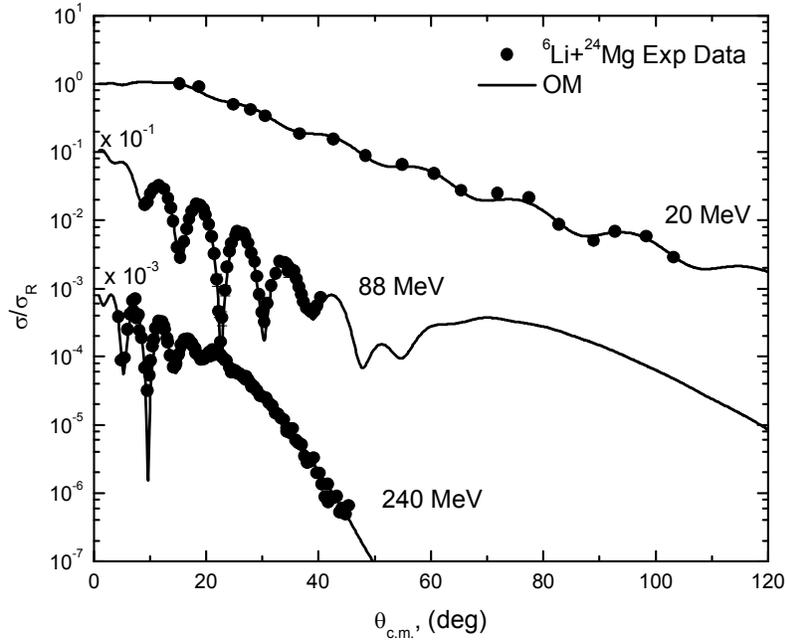

**Fig. 3:** Elastic scattering $^6$Li + $^{24}$Mg experimental ADs (solid circles) versus OM calculations (solid line) at $E_{lab}$ = 20, 88, and 240 MeV.

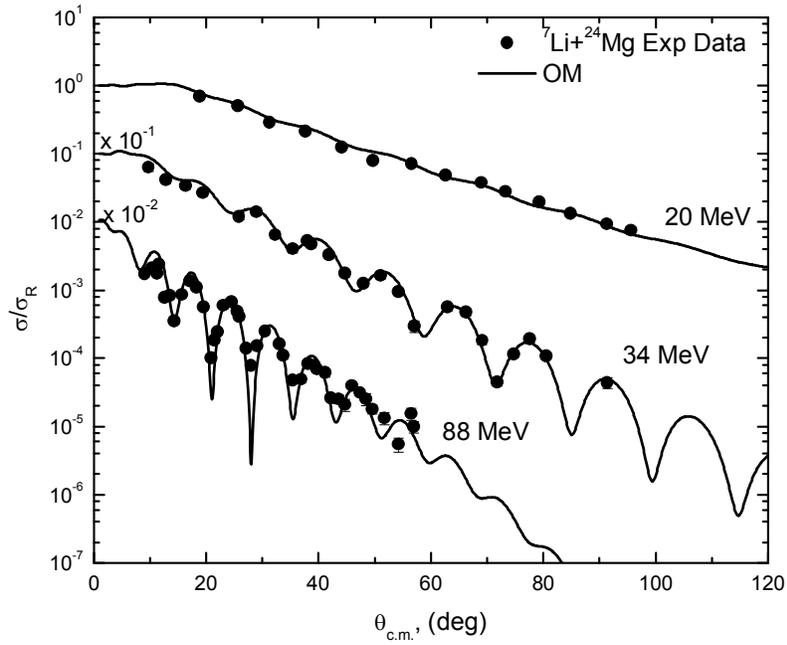

**Fig. 4:** Elastic scattering $^7$Li + $^{24}$Mg experimental ADs (solid circles) versus OM calculations (solid line) at $E_{lab}$ = 20, 34, and 88 MeV.

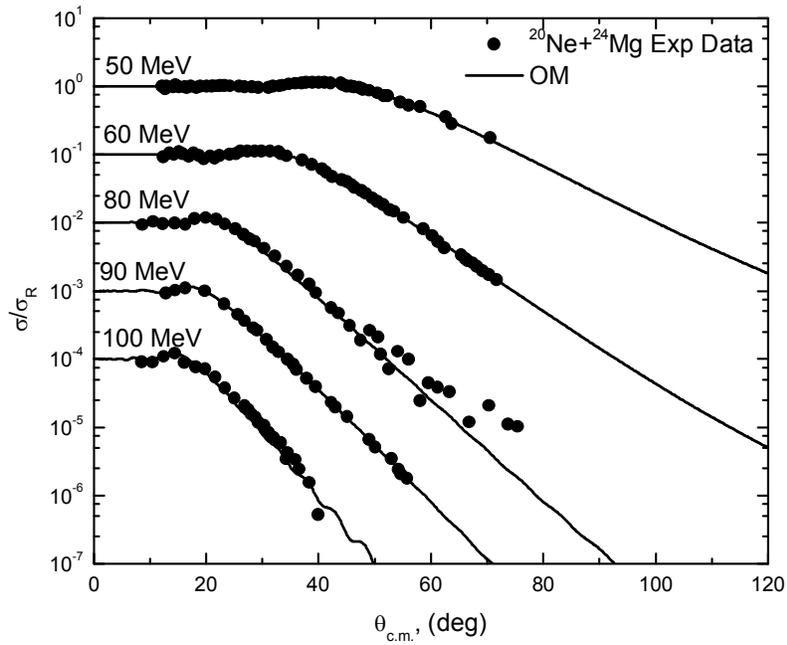

**Fig. 5:** Elastic scattering $^{20}$Ne + $^{24}$Mg experimental ADs (solid circles) versus OM calculations (solid line) at $E_{lab}$ = 50, 60, 80, 90, and 100 MeV.

As shown in table I, the $^{20}$Ne + $^{24}$Mg ADs are well fitted using a very shallow real potential depth, in contrast to different nuclear systems induced by $^{20}$Ne, even by lighter target such as $^{16}$O and $^{12}$C. The optimal extracted imaginary potential depth at the different considered energies is nearly compatible with the neighboring nuclear systems such as $^{20}$Ne + $^{16}$O and $^{20}$Ne + $^{20}$Ne [7-9]. The experimental ADs for the $^{20}$Ne + $^{24}$Mg system "see Fig. 5" showed a Coulomb nuclear interference peak. The position of this peak is found to be shifted toward smaller scattering angles with increasing the bombarding energy.

**Table I:** Optimal OMP parameters for $^6$Li, $^7$Li, $^{20}$Ne + $^{24}$Mg systems.

| $E_{lab}$ (MeV) | $V_0$ (MeV) | $r_V$ (fm) | $a_V$ (fm) | $W_0$ (MeV) | $r_W$ (fm) | $a_W$ (fm) | $\chi^2/N$ | $\sigma$ (mb) |
|---|---|---|---|---|---|---|---|---|
| $^6$Li + $^{24}$Mg | | | | | | | | |
| 20 | 182.15 | 1.079 | 0.905 | 19.64 | 1.867 | 0.764 | 1.6 | 1266 |
| 88 | 166.14 | 1.079 | 0.893 | 20.21 | 1.867 | 0.844 | 8.5 | 1703 |
| 240 | 124.96 | 1.079 | 0.95 | 19.32 | 1.867 | 0.95 | 2.7 | 1666 |
| $^7$Li + $^{24}$Mg | | | | | | | | |
| 20 | 119.72 | 1.286 | 0.853 | 45.0 | 1.739 | 0.809 | 2.5 | 1522 |
| 34 | 109.92 | 1.286 | 0.853 | 30.95 | 1.739 | 0.809 | 2.51 | 1624 |
| 88 | 118.13 | 1.286 | 0.853 | 36.4 | 1.739 | 0.809 | 18.7 | 1740 |
| $^{20}$Ne + $^{24}$Mg | | | | | | | | |
| 50 | 17.02 | 1.25 | 0.65 | 17.42 | 1.25 | 0.65 | 0.11 | 1088 |
| 60 | 20.79 | 1.25 | 0.65 | 20.36 | 1.25 | 0.65 | 0.19 | 1272 |
| 80 | 25.22 | 1.25 | 0.65 | 24.72 | 1.25 | 0.65 | 5.8 | 1652 |
| 90 | 19.47 | 1.25 | 0.65 | 20.89 | 1.25 | 0.65 | 0.17 | 1690 |
| 100 | 30.2 | 1.25 | 0.65 | 21.43 | 1.25 | 0.65 | 4.9 | 1809 |

### B. Analysis of $^{6,7}$Li and $^{20}$Ne + $^{24}$Mg systems using SPP

The considered data are microscopically analyzed using real SPP in addition to an imaginary potential taken as a factor times the real SPP, the approach namely, (Real SPP + Imag. SPP). As shown in Figs. 6–8, the comparison between the experimental $^{6,7}$Li + $^{24}$Mg and $^{20}$Ne + $^{24}$Mg elastic scattering ADs and theoretical calculations using the (Real SPP + Imag. SPP) approach is fairly good. Two adjustable parameters, $N_{RSPP}$ and $N_{ISPP}$, which are the renormalization factors for the real and imaginary SPP, respectively, were used to fit the considered data. The optimal extracted parameters using this approach are listed in table II. The implemented nuclear potential has the following form:

$$U(R) = V_C(R) - N_{RSPP} V^{DF}(R) - i N_{ISPP} V^{DF}(R) \tag{10}$$

The average extracted $N_{RSPP}$ and $N_{ISPP}$ values are 0.701±0.118 and 0.603±0.175, respectively, for $^6$Li + $^{24}$Mg system, and for $^7$Li + $^{24}$Mg system, the average extracted $N_{RSPP}$ and $N_{ISPP}$ values are 0.681±0.328 and 0.653±0.204, respectively. These results show the need to reduce real SPP strength by ~ 30-32% in order to describe the data. On the other hand, for the $^{20}$Ne + $^{24}$Mg system induced by the $^{20}$Ne nucleus, which is more bounded than $^{6,7}$Li, the average extracted $N_{RSPP}$ and $N_{ISPP}$ values are 0.87±0.18 and 0.75±0.1, respectively. The lower renormaliztion factor for the real potential strength indicates the weaker binding nature. Consequently, the extracted $N_{RSPP}$ values within the (Real SPP + Imag. SPP) approach emphasize the weak binding nature of $^{6,7}$Li projectiles in comparison with $^{20}$Ne nucleus.

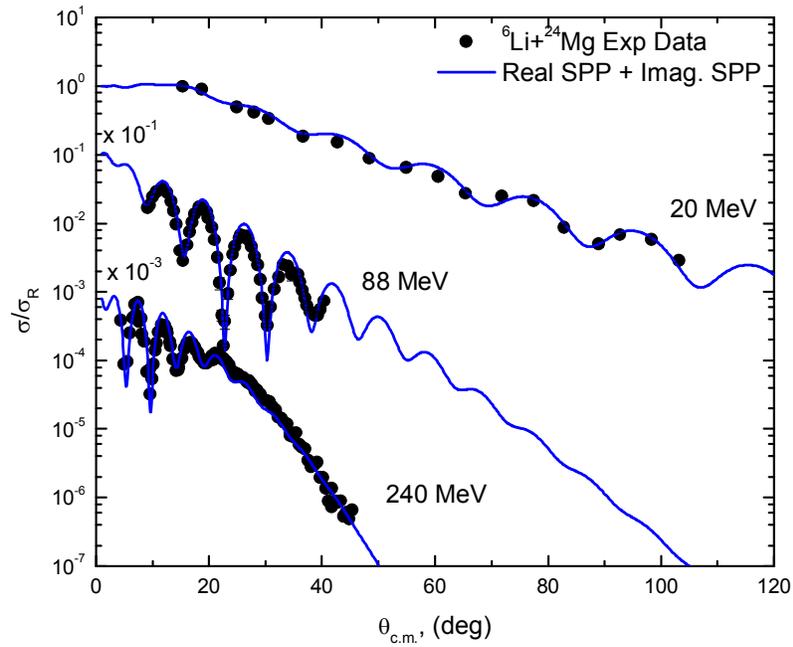

**Fig. 6:** Elastic scattering $^6$Li + $^{24}$Mg experimental ADs (solid circles) versus theoretical calculations using the (Real SPP + Imag. SPP) approach (solid line) at $E_{lab}$= 20, 88, and 240 MeV.

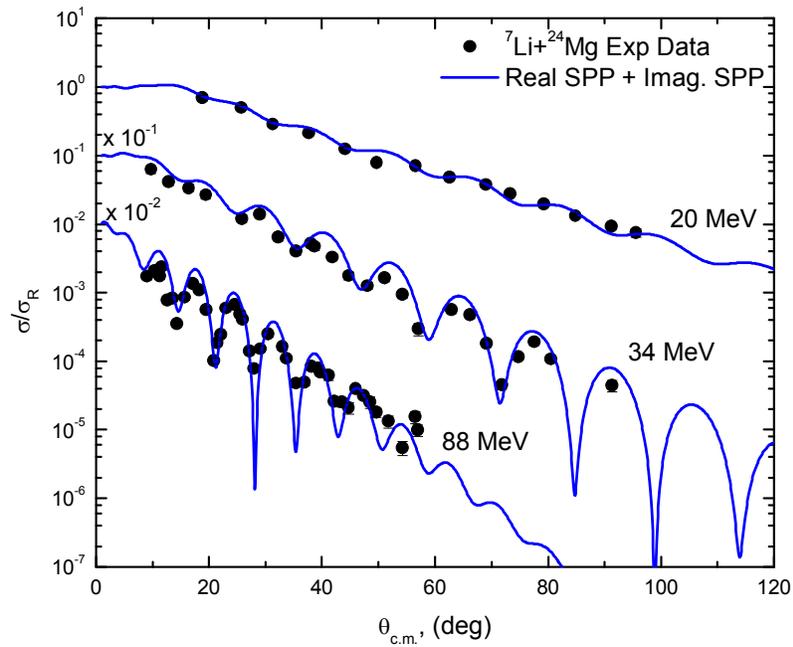

**Fig. 7:** Elastic scattering $^7$Li + $^{24}$Mg experimental ADs (solid circles) versus theoretical calculations using the (Real SPP + Imag. SPP) approach (solid line) at $E_{lab}$= 20, 34, and 88 MeV.

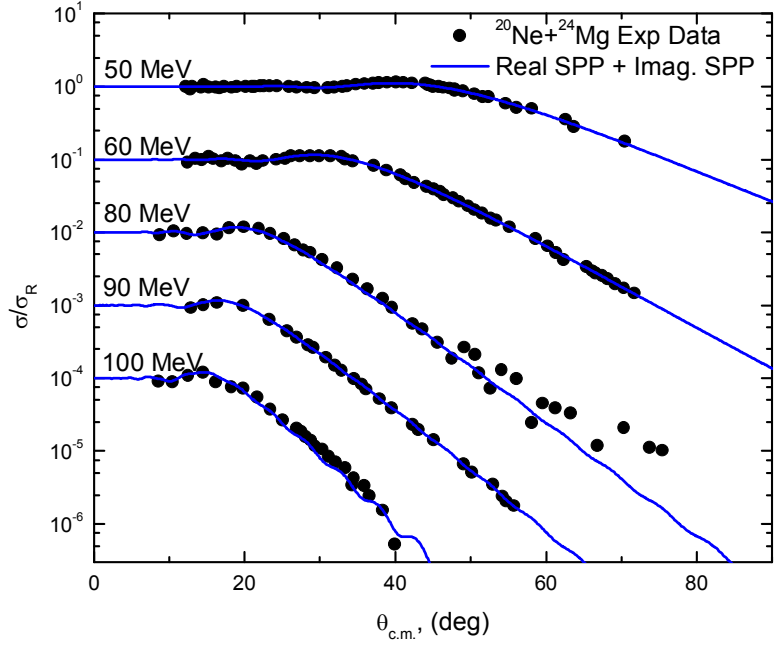

**Fig. 8:** Elastic scattering $^{20}$Ne + $^{24}$Mg experimental ADs (solid circles) versus theoretical calculations using the (Real SPP + Imag. SPP) approach (solid line) at $E_{lab}$= 50, 60, 80, 90, and 100 MeV.

**Table II:** Optimal potential parameters for $^6$Li, $^7$Li, $^{20}$Ne + $^{24}$Mg nuclear systems extracted from analysis using the (Real SPP + Imag. SPP) approach.

| $E_{lab}$ (MeV) | $N_{RSPP}$ | $N_{ISPP}$ | $\chi^2/N$ | $\sigma$ (mb) |
|---|---|---|---|---|
| $^6$Li + $^{24}$Mg | | | | |
| 20 | 0.725 | 0.409 | 5.6 | 1274 |
| 88 | 0.572 | 0.649 | 32.2 | 1662 |
| 240 | 0.806 | 0.75 | 7.1 | 1649 |
| $^7$Li + $^{24}$Mg | | | | |
| 20 | 1.06 | 0.87 | 4.1 | 1534 |
| 34 | 0.495 | 0.464 | 17.3 | 1612 |
| 88 | 0.487 | 0.626 | 35.5 | 1749 |
| $^{20}$Ne + $^{24}$Mg | | | | |
| 50 | 0.652 | 0.648 | 0.1 | 950.0 |
| 60 | 0.816 | 0.741 | 0.18 | 1262 |
| 80 | 0.989 | 0.888 | 5.6 | 1635 |
| 90 | 0.788 | 0.796 | 0.15 | 1692 |
| 100 | 1.102 | 0.683 | 4.0 | 1762 |

## C. Analysis of $^{6,7}$Li and $^{20}$Ne + $^{24}$Mg systems using CFP

Within the framework of the more microscopic CFM, the considered data is analyzed microscopically using real and imaginary CFPs prepared as described in detail in the previous section. The implemented CFPs to describe the ADs data for the $^6$Li + $^{24}$Mg, $^7$Li + $^{24}$Mg, and $^{20}$Ne + $^{24}$Mg system are presented in Fig. 2. As shown in Figs. 9–11, the comparisons between the experimental elastic scattering ADs and CFM calculations are fairly good for the considered systems, with the optimal extracted parameters listed in table III. In this case, the data is fitted using two parameters, $N_{RCF}$ and $N_{ICF}$, which are the renormalization factors for the real and imaginary CFPs, respectively. This approach is denoted as (Real CFP + Imag. CFP). The implemented potential has the form:

$$U(R) = V_C(R) - N_{RCF} V^{CF}(R) - i N_{ICF} V^{CF}(R) \qquad (11)$$

The average extracted $N_{RCF}$ and $N_{ICF}$ values are 0.58±0.14 and 1.17±0.07, respectively, for the $^6$Li + $^{24}$Mg system. While, for the $^7$Li + $^{24}$Mg system, the average extracted $N_{RCF}$ and $N_{ICF}$ values are 0.68±0.27 and 1.26±0.56, respectively. These results show the need to reduce real CFP strength by ~ 32–42% in order to describe the data. Similar finding was obtained from the CFM analysis for the $^{20}$Ne + $^{24}$Mg system, the average extracted $N_{RCF}$ and $N_{ICF}$ values are 0.61±0.08 and 1.46±0.26, respectively. The required reduction in the real CFP strength for the $^6$Li + $^{24}$Mg, $^7$Li + $^{24}$Mg and $^{20}$Ne + $^{24}$Mg systems is close to each other.

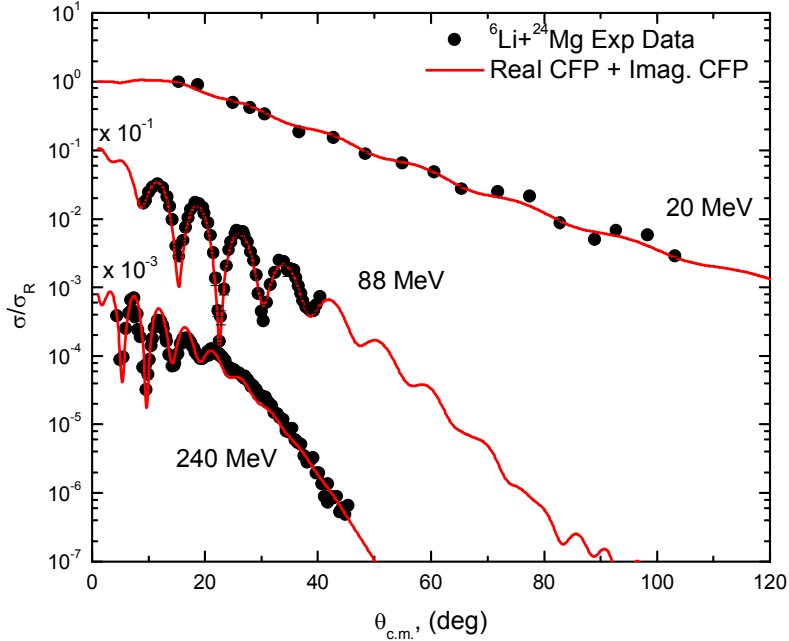

**Fig. 9:** Elastic scattering $^6$Li + $^{24}$Mg experimental ADs (solid circles) versus theoretical calculations using the (Real CFP + Imag. CFP) approach (solid line) at $E_{lab}$= 20, 88, and 240 MeV.

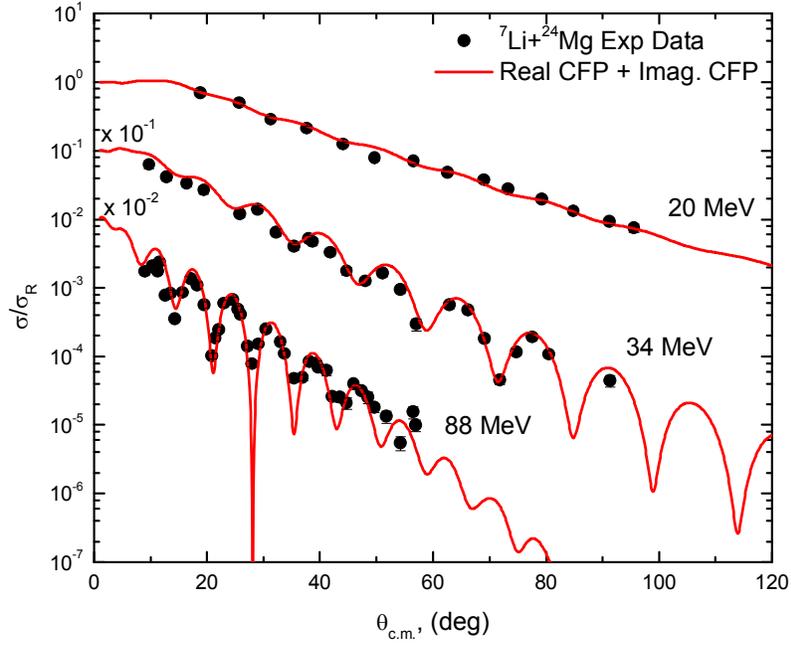

**Fig. 10:** Elastic scattering $^{7}$Li + $^{24}$Mg experimental ADs (solid circles) versus theoretical calculations using the (Real CFP + Imag. CFP) approach (solid line) at $E_{lab}$= 20, 34, and 88 MeV.

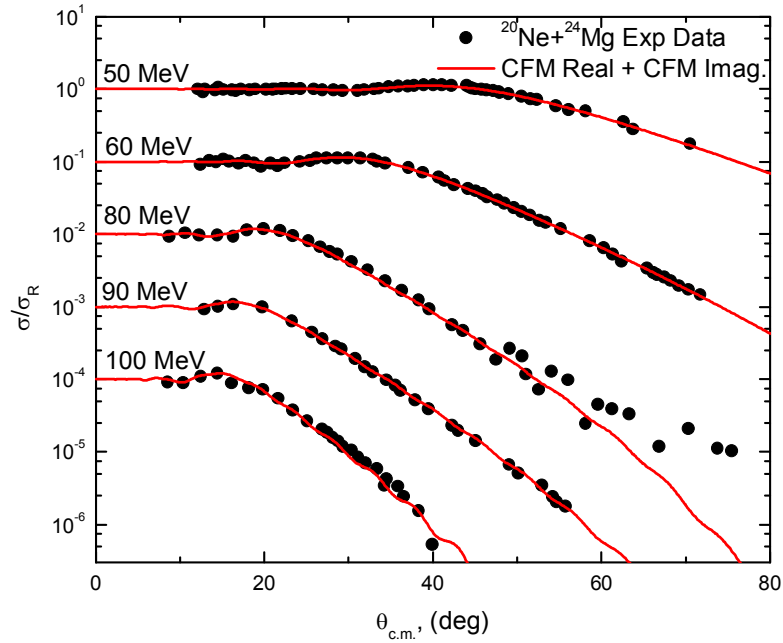

**Fig. 11:** Elastic scattering $^{20}$Ne + $^{24}$Mg experimental ADs (solid circles) versus theoretical calculations using the (Real CFP + Imag. CFP) approach (solid line) at $E_{lab}$= 50, 60, 80, 90, and 100 MeV.

**Table III:** Optimal potential parameters for $^6$Li, $^7$Li, $^{20}$Ne + $^{24}$Mg nuclear systems extracted from analysis using the (Real CFP + Imag. CFP) approach.

| $E_{lab}$ (MeV) | $N_{RCF}$ | $N_{ICF}$ | $\chi^2/N$ | $\sigma$ (mb) |
|---|---|---|---|---|
| $^6$Li + $^{24}$Mg | | | | |
| 20 | 0.454 | 1.22 | 3.8 | 1253 |
| 88 | 0.559 | 1.09 | 10.3 | 1691 |
| 240 | 0.727 | 1.19 | 9.1 | 1607 |
| $^7$Li + $^{24}$Mg | | | | |
| 20 | 0.998 | 1.9 | 2.48 | 1713 |
| 34 | 0.529 | 0.846 | 6.0 | 1634 |
| 88 | 0.526 | 1.046 | 22.5 | 1763 |
| $^{20}$Ne + $^{24}$Mg | | | | |
| 50 | 0.48 | 1.28 | 0.09 | 946.6 |
| 60 | 0.60 | 1.49 | 0.19 | 1264 |
| 80 | 0.69 | 1.71 | 6.1 | 1625 |
| 90 | 0.61 | 1.71 | 0.39 | 1718 |
| 100 | 0.65 | 1.12 | 2.9 | 1713 |

The reaction cross section values ($\sigma_R$) extracted from the performed analysis utilizing the OMP, (Real SPP + Imag. SPP) and (Real CFP + Imag. CFP) approaches for the $^6$Li + $^{24}$Mg, $^7$Li + $^{24}$Mg and $^{20}$Ne + $^{24}$Mg systems are listed in tables I, II and II, respectively. The energy dependence on $\sigma_R$ values is plotted as shown in Fig. 12. For each system, the extracted $\sigma_R$ values within the framework of the implemented approaches are close to each other, and they increase with increasing energy. The following polynomial functions can be used to express this dependence:

- in the case of the $^6$Li + $^{24}$Mg system: $\sigma_R(E) = 1088.6 + 9.37\,E - 0.03\,E^2$
- in the case of the $^7$Li + $^{24}$Mg system: $\sigma_R(E) = 1331.6 + 10.98\,E - 0.07\,E^2$
- in the case of the $^{20}$Ne + $^{24}$Mg system: $\sigma_R(E) = -1216.18 + 58.82\,E - 0.29\,E^2$

In general, the ADs for stable $^{20}$Ne elastically scattered from $^{24}$Mg target at energies 50–100 MeV exhibit the classical Fresnel diffraction scattering pattern as shown in Fig. 5. On the other hand, the ADs for processes induced by weakly bound nuclei such as $^{6,7}$Li deviate significantly from the oscillatory pattern. This deviation from the oscillatory pattern was also previously observed in the scattering of $^{11}$Be (one neutron halo nucleus) [37, 38] from different targets such as $^{64}$Zn [39, 40], $^{120}$Sn [41], $^{197}$Au [42], and $^{209}$Bi [43, 44] which exhibit a strong suppression of the Fresnel peak due to the break-up effects. The extracted $N_{ICF}$ values from the analysis of $^{6,7}$Li + $^{24}$Mg systems showed an increase at the lowest energy, which gives an evidence for the absence of the usual threshold anomaly (TA).

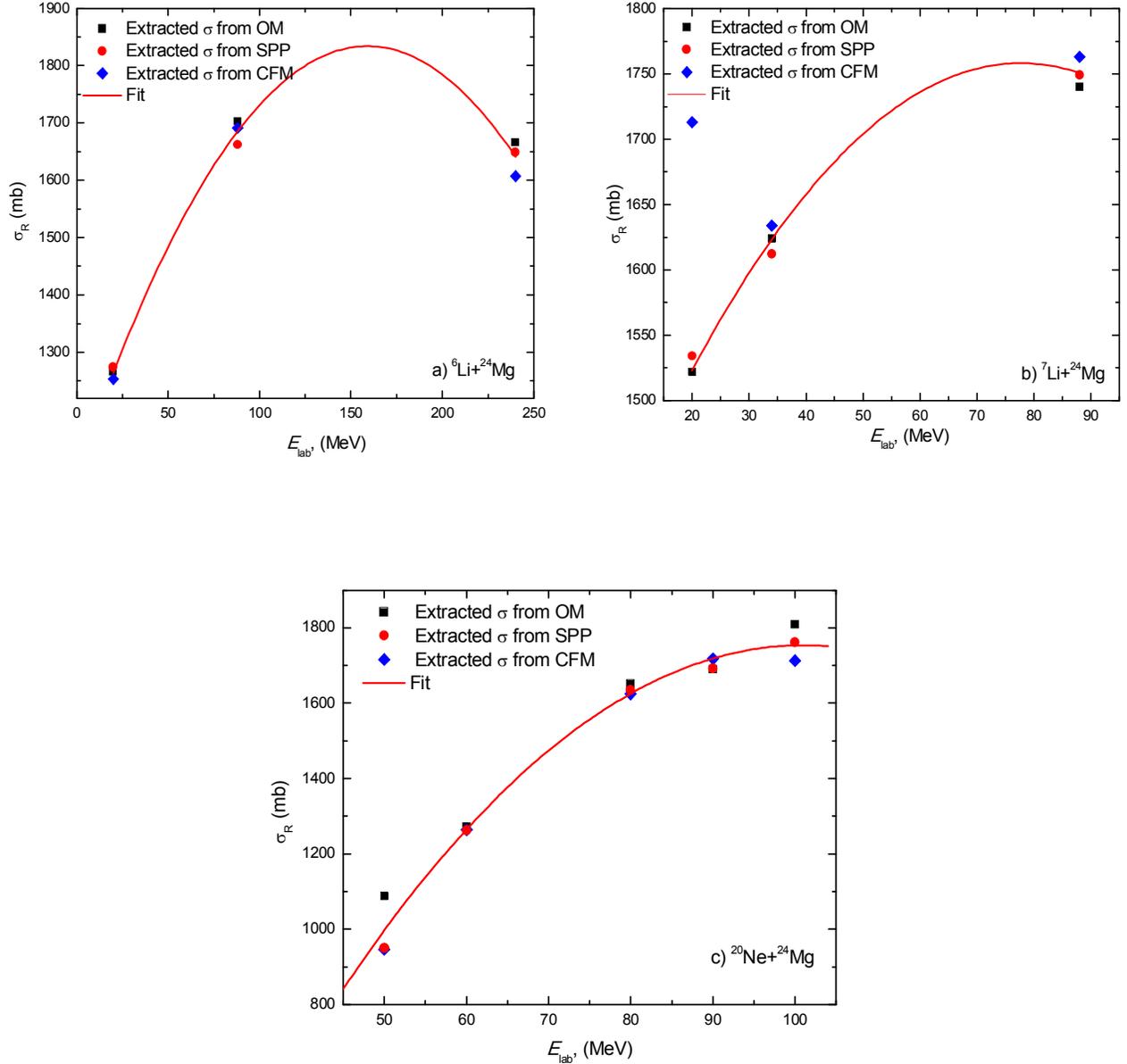

**Fig. 12:** Extracted $\sigma_R$ values from OM, SPP and CFM calculations versus energy for a) $^6$Li + $^{24}$Mg system, b) $^7$Li + $^{24}$Mg system, and c) $^{20}$Ne + $^{24}$Mg system

### IV. SUMMARY

The $\alpha + ^{16}$O model of the $^{20}$Ne nucleus has recently been applied to investigate various systems, including the $^{20}$Ne nucleus (either as a projectile or a target) with a significant success, among these systems: $^{20}$Ne + $^{20}$Ne, $^{20}$Ne + $^{16}$O, $\alpha + ^{20}$Ne, and p + $^{20}$Ne systems [7-11]. Unfortunately, the only performed experimental measurements for $^{20}$Ne + $^{24}$Mg system was done many years ago [17], where the measured $^{20}$Ne + $^{24}$Mg ADs at $E_{lab}$ = 50, 60 80, 90 and 100 MeV were analyzed from the phenomenological perspective. In the current study, these data is subjected to detailed analysis using various approaches OM, SPP and CFM. Additionally, the elastic scattering ADs for $^6$Li + $^{24}$Mg at $E_{lab}$ = 20, 88, and 240 MeV and $^7$Li + $^{24}$Mg system at $E_{lab}$ = 20, 34, and 88 MeV are also studied. Analyses within the aforementioned approaches give satisfactory descriptions for the considered data. The following findings are drawn:

- the possibility of the appearance of the $\alpha$ + $^{16}$O structure in the ground state of $^{20}$Ne nucleus.
- The extracted $N_{RSPP}$ values from the performed analysis within the approach (Real SPP + Imag. SPP) reflect the weak binding nature of $^{6,7}$Li projectiles in comparison with $^{20}$Ne nucleus.
- The energy dependence on the renormalization factors for real and imaginary SPP as well as on those for CFPs revealed that $^6$Li + $^{24}$Mg and $^7$Li + $^{24}$Mg systems exhibit BTA, which was widely observed in different systems induced by weakly projectiles.

**Acknowledgments**
Sh. Hamada is funded by a full postdoctoral scholarship from the Ministry of Higher Education of the Arab Republic of Egypt.

**References**
1. H. Morinaga (1956) *Phys. Rev.* 101: 254.
2. K. Ikeda, N. Tagikawa, and H. Horiuchi, *Prog. Theo. Phys. Suppl.* 464. (1968).
3. M. Freer (2007). The clustered nucleus—cluster structures in stable and unstable nuclei *R*eports on Progress in Physics 70: 2149.
4. F. Hoyle, D. N. F. Dunbar, and W. A. Wenzel, Phys. Rev. 92 (1953)
5. C. Cook, W. A. Fowler and T. Lauritsen *Phys. Rev.* 107 (1957) 508
6. G. R. Satchler, Direct Nuclear Reactions, Clareton, Oxford, 1983
7. Yongxu Yang, Xinrui Zhang and Qingrun Li, J. Phys. G: Nucl. Part. Phys. 42 (2015) 015101
8. Sh. Hamada, N. Keeley, K. W. Kemper, K. Rusek, Phys. Rev. C 97 (2018) 054609
9. Sh. Hamada et al., Braz. J. Phys. 51 (2021) 780.
10. Yong-Xu Yang, Hai-Lan Tan, and Qing-Run Li, Phys. Rev. C 82 (2010) 024607
11. P.-T. Ong, Y.-X. Yang, and Q.-R. Li, Eur. Phys. J 41 (2009) 229
12. K. Bethge, C. M. Fou, and R. W Zurmuhle, Nucl. Phys. A123 (1969) 521
13. C. B. Fulmer et al., Nucl. Phys. A356 (1981) 235
14. X. Chen, Y.- W. Lui, H. L. Clark, Y. Tokimoto, and D. H. Youngblood, Phys. Rev. C 80 (2009) 014312
15. G.E.Moore, K.W. Kemper, L.A. Charlton, Phys. Rev. C11 (1975) 1099
16. M.F. Steeden et al., J. Phys. G 6 (1980) 501
17. P. Belery, Th. Delbar, Gh. Gregoire, K. Grotowski, and N. S. Wall, Phys. Rev. C 23 (1981) 2503.
18. I. J. Thompson, Comput. Phys. Rep.7 (1988) 167
19. J. Cook, Nucl. Phys. A 388 (1982) 153.
20. B. V. Carlson and D. Hirata, Phys. Rev. C62 (2000) 054310
21. L. C. Chamon, B. V. Carlson and L. R. Gasques, Comp. Phys. Comm. 267 (2021) 108061
22. F. Hinterberger et al., Nucl. Phys. A111 (1968) 265.
23. L. Bimbot et al., Nucl. Phys. A210 (1973) 397.
24. J. B. A. England et al., Nucl. Phys. A475 (1987) 422.
25. H. H. Duhm, Nucl. Phys. A118 (1968) 563.

26. J. Lega and P. C. Macq, Nucl. Phys. A218 (1974) 429.
27. S. L. Tabor, D. F. Geesaman, W. Henning, D. G. Kovar, and K. E. Rehmt, Phys. Rev. C17 (1978) 2136.
28. B. Buck, J. C. Johnston, A. C. Merchant, and S. M. Perez, Phys. Rev. C 52 (1995 ) 1840.
29. Sh. Hamada and Awad A. Ibraheem, Int. J. Mod. Phys. E28 (2019) 1950108.
30. Sh. Hamada et al., Revista Mexicana de Fisica 66(3) (2020) 322.
31. Sh. Hamada and Awad A. Ibraheem, Revista Mexicana de Fisica 67 (2) (2021) 276.
32. Sh. Hamada, Norah A. M. Alsaif and Awad A. Ibraheem, Phys. Scr. 96 (2021) 055306.
33. Sh. Hamada and Awad A. Ibraheem, Braz. J. Phys. 52 (2022) 29.
34. Sh. Hamada, N. Burtebayev and Awad A. Ibraheem, Revista Mexicana de Fisica 68 (2022) 031201.
35. Sh. Hamada and A. A. Ibraheem, Journal of Taibah University for Science 16 (2022) 163.
36. Sh. Hamada and Awad A. Ibraheem, Int. J. Mod. Phys. E31, 3 (2022) 2250019.
37. I. Tanihata et al., Phys. Lett. B206 (4) (1988) 592
38. T. Aumann et al., Phys. Rev. Lett. 84 (2000) 35.
39. A. Di Pietro et al., Phys. Rev. Lett. 105 (2010) 022701.
40. A. Di Pietro et al., Phys. Rev. C. 85 (2012) 054607.
41. L. Acosta et al., Eur. Phys. J. A 42 (3) (2009) 461.
42. V. Pesudo et al., Phys. Rev. Lett. 118 (2017) 152502.
43. M. Mazzocco et al., Eur. Phys. J. A 28 (3) (2006) 295.
44. M. Mazzocco et al., Eur. Phys. J. Spec. Top. 150 (1) (2007) 37